# Runtime Reconfiguration of J2EE Applications

**Jasminka Matevska-Meyer, Sascha Olliges, Wilhelm Hasselbring**

*Department of Computing Science, Software Engineering Group, University of Oldenburg, 26121 Oldenburg, Germany*
*matevska-meyer@informatik.uni-oldenburg.de*
*olliges@informatik.uni-oldenburg.de*
*hasselbring@informatik.uni-oldenburg.de*

ABSTRACT: *Runtime reconfiguration considered as "applying required changes to a running system" plays an important role for providing high availability not only of safety- and mission-critical systems, but also for commercial web-applications offering professional services. Hereby, the main concerns are maintaining the consistency of the running system during reconfiguration and minimizing its down-time caused by the reconfiguration.*

*This paper focuses on the platform independent subsystem that realises deployment and redeployment of J2EE modules based on the new J2EE Deployment API as a part of the implementation of our proposed system architecture enabling runtime reconfiguration of component-based systems. Our "controlled runtime redeployment" comprises an extension of hot deployment and dynamic reloading, complemented by allowing for structural changes.*

## 1    Introduction

The permanent change of requirements on software systems necessitates their evolution. Reengineering approaches aim at a reasonable re-design of the system and variability management approaches concentrate on designing systems which include variation points for post-deployment system adaptation. Runtime reconfiguration as *applying required changes to a running system* plays an important role for providing high availability of software systems. Main issues are maintaining the consistency of the system during runtime reconfiguration and minimising its down-time caused by the reconfiguration. Due to that, techniques are required which determine the parts of the system to be halted during reconfiguration, and, accordingly, the parts of the system which can continue execution during reconfiguration [MAT 03:2]. In order to identify those parts as a minimal set of affected components, we need a system description, which provides an information of its runtime behaviour basically concerning uses dependencies among instances of components [MAT 03:3]. Furthermore, we have to be able to re-compose the system during its runtime.

Our approach to runtime reconfiguration concerning the deployment called *controlled runtime redeployment* presents an extension of the concepts of hot deployment and dynamic reloading [IBM 03]. Additionally, we consider consistent structural changes of the running system [MAT 03:2].

This paper is organised as follows. First, we briefly present our approach to enabling reconfiguration of component-based systems at runtime (Section 3); next, we propose a system architecture (Figure 1). In Section 2.1 we present our implementation of the J2EE Deployment API [OLL 04]. Finally, we conclude and indicate further work in Section 3.

## 2    Enabling Reconfiguration of Component-Based Systems at Runtime

We aim at reconfiguration of component-based systems at runtime as *applying required changes to a running system*. We distinguish between three different types of reconfiguration according to their reconfiguration effort: (1) functional, (2) non-functional, and (3) structural. All types of reconfiguration can occur on different levels of granularity (i.e., can address the entire system or a single sub-component). *Functional reconfigurations* include changes to the functionality of a single component as well as of a particular subsystem, even of the entire system. *Non-functional reconfigurations* are concerned with the quality of service (QoS) of the system and can affect single components (sub-systems) or the architecture. *Structural reconfigurations* consider both, changing the interface of a single component and changing dependencies among components (architectural changes of a system).

We observe a running system at a particular time interval from receiving a reconfiguration request until reconfiguration completion. In an already deployed and running system we determine time-constrained use dependencies among instances of components, which are constrained by specified structural dependencies and specified or derived component protocol information. Knowing the current state of all possibly affected components, and their future behaviour, we can exclude past dependencies and late future ones. This allows us to build a minimal runtime dependency graph matching the particular reconfiguration request [MAT 03:2].

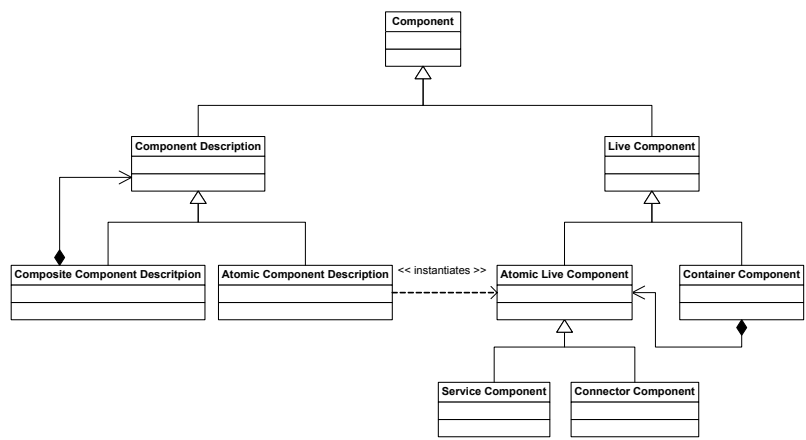

**Figure 1.** *C3-Meta Model*

Our meta-model in Figure 1, described in [MAT 03:3] presents a Composite [GAM 95] structural static and runtime view of the system, thus enabling hierarchical decomposition of the system behaviour and its consistency check and composition after reconfiguration.

For this paper container components are the essential extension to establish modelling of deployment and runtime properties of the system. A container provides the runtime environment for the live components [OMG 03, SUN 03] To describe the runtime behaviour of the system we use *service effect automata* (as specified in [REU 03]). For determining the reconfiguration point in time we propose a special extension of message sequence charts called *live sequence charts* [DAM 01] because they can express liveliness and timing constraints.

Finally, applying required changes to an already deployed and running system usually triggers changes in a system configuration and implies its reconstruction and redeployment to obtain a consistent system after reconfiguration. A major problem to be solved here is managing runtime dependencies among the components. Our concept of *controlled runtime redeployment* presents an extension of the concepts of hot deployment and dynamical reloading [IBM 03]. We additionally allow structural changes of the running system [MAT 03:2] and manage consistency problems in contrast to both other concepts mentioned before, which only allow a simple swap of an application or a single component at runtime.

### 2.1 *Implementation of the J2EE Deployment API*

Hot redeployment of software components in the J2EE (Java 2 Enterprise Edition) platform is implemented by the J2EE product providers as an optional feature for component developers who continuously need to execute tests in a running environment. As a consequence, hot redeployment was and actually is an operation that potentially invalidates existing user sessions. This is no problem for developers, as loosing the session state is not critical while debugging and testing components. In productive systems, that expose their services to real users, deployment is a time- and error-sensitive process. While program correctness is only partially affected by the deployment process, time is an important factor when dealing with component deployment. In most cases, maintenance downtime of productive systems is considered a big problem because other business processes depend on the systems availability.

The Deployment API specification [SEA 03] introduced as a part of the new J2EE 1.4 specification [SUN 03] takes the concept of redeployment one step further by specifying redeployment to be transparent to users thus allowing it to be used in productive systems. The possibility to perform configuration changes in a running system without invalidating running sessions will significantly reduce its downtime.

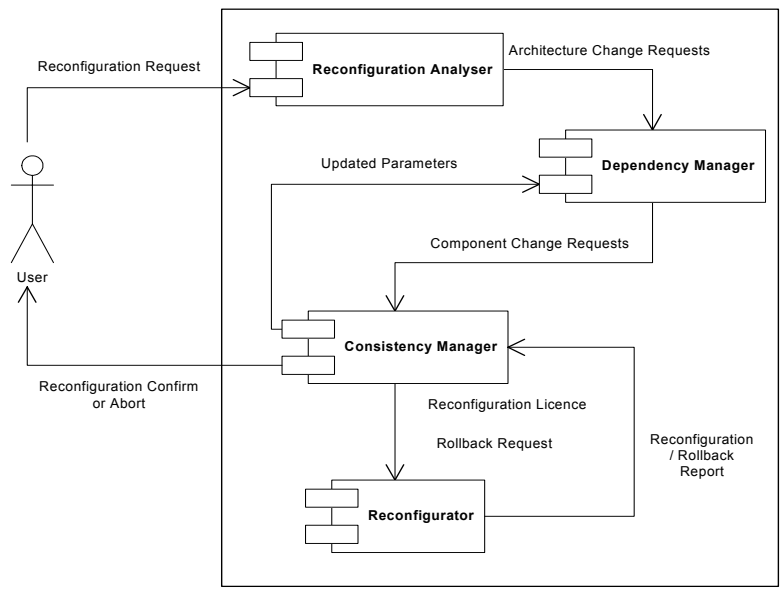

**Figure 2.** *Reconfiguration Manager*

However, no J2EE application server includes a working implementation of the Deployment API specification yet. Our project "J2EE Deployment API Implementation" [OLL 04] developed the technical basis for an API implementation for the JBoss Application Server. We now work on implementing the redeployment functionality which is marked optional in the specification. The resulting API implementation will then be integrated into a complete system enabling configuration, reconfiguration, deployment and redeployment of J2EE applications in production systems.

We name our system PIRMA (*Platform Independent Reconfiguration Manager*). It is activated on *reconfiguration requests*. It consists of the following four top-level components [MAT 03]: *Reconfiguration Analyser*, *Dependency Manager*, *Consistency Manager* and *Reconfigurator*. Our implementation of the J2EE Deployment API covers the fundamental functionality of the reconfigurator and provides an interface to the consistency manager.

*2.1.1 Basic concepts of the Redeployment Implementation*

Our project exploits the potentials of the JBoss interceptor stack technology [STA 02] to support redeployment. In JBoss server each J2EE component is deployed inside a manageable container component. That container is configured with a chain of interceptor objects that handle the configurable system-level aspects of an EJB component which are: transaction demarcation, persistence, authentication, authorization and remote communication as well as instance pooling and optionally clustering. Other responsibilities of interceptors in these chains are invocation routing, logging and as a matter of particular interest an interceptor that manages container shutdown operations. The *CleanShutdownInterceptor's* responsibility is to wait for the completion of running invocations on the component container it is configured for and to deny further invocations. We plan to use a similar mechanism to support transparent redeployment operations. A newly introduced interceptor will let any outstanding invocation that started a new transaction complete while new invocations wait on some synchronization barrier. After invocation completion the component hosted in the container will be replaced with a new version. Unlike the *CleanShutdownInterceptor* which ignores the transaction attribute of the operation to be invoked, the new interceptor has to intercept operations that are configured to start new transactions. Thus transactions are guaranteed to complete and no configuration change is performed while a transaction is running. Needless to say that there are some restrictions on the configuration changes allowed in such redeployments. The Deployment API specification states that the runtime configuration must remain the same for a J2EE module to be successfully redeployed. However, this restriction may be weakened to some extent. It should be possible to support structural changes to module parts that conform to certain requirements. The component type determines the criteria the component has to satisfy and therefore plays a key role in the distinction of whether the component is safe to structural change or not.

**Session bean** type EJB components are by definition extensions of the client that created them. The EJB specification [SUN 03:2] defines two types of such components. A stateful session bean instance contains conversational state that must be retained across methods and transactions. The session bean container sometimes needs to transfer the state of the hosted beans to secondary storage for performance reasons. This transfer is called passivation. The operation of bringing beans back to life is called activation. To support this, the interface implemented by the session bean types contains call-back methods that the container invokes to inform a component about its passivation or activation. It's the instances responsibility to ensure, that upon return from a call to the passivate method its fields are ready to be stored via Java serialization. As serialization depends on the serialized type's structure, a stateful session bean is not safe to structural change. The stateless session bean component type contains no conversational state between method invocations. Therefore bean instances of this type are interchangeable. As there is no need to preserve state information when switching versions of the component this would be no problem to redeployment. On the other hand, as session beans are client extensions, their structure is incorporated into the client itself and therefore a stateless session bean may only be redeployed if it is not referenced by (unchanged) remote clients. This is always the case when dealing with local EJB components (see chapter 6.5 of the EJB specification [SUN 03:2]).

An **entity bean** type EJB component is an object-oriented view of information entities, like a person or an account for example, that are stored in a database or an existing enterprise application. As an entity bean is a view on data located elsewhere, it contains no state. Depending on the container's configuration it may be cached in the application server, but under any circumstances it is guaranteed that modifications to the entity are written to the data store when the current transaction is committed. The real problem that occurs when dealing with redeployments of entity bean components is that when its structure changes, the data structures in the associated persistence store most of the time needs to be changed, too. While this is surely possible, such an operation may be a long-running task, which is not acceptable in redeployments that should be transparent to users of the system. A possible solution to this would be using a second data store somehow externally synchronizing its contents with the first. That second data store then could be used for persistence of the entities new version. The switch of the data store is performed by configuring a new resource manager for the new version of the entity. Again it is not possible to change the structure of the EJB component if it is referenced by (unchanged) remote clients.

**Message driven** bean type components have no client-visible identity. A message driven bean contains no conversational state specific to a client, but they of course may contain instance variables that constitute state valid across the handling of client messages. However, the EJB specification [SUN 03:2] states that all instances of a message driven bean are equivalent and therefore a message may be send to any instance. As a message driven bean is by definition an asynchronous message receiver, a redeployment operation may temporarily disable message routing and exchange the associated bean.

The above observations now can be summarized: The distinction of whether an EJB component is safe to structural change (i.e. interface modification) or not can be boiled down to the question whether (unchanged) remote clients hold references to it or not and if it contains conversational state or not. To detect if a component is safe to configuration change (or even removal) another interceptor may be introduced, this time intercepting invocations of the components home interface (EJBHome) which is used to create, find and remove handles (stubs) to EJB objects.

*2.1.2    Related Work*

The JBoss open source project itself started to work on an implementation of the Deployment API. Being a new project, the current sources are incomplete and do not (yet) include any support for redeployment. Anyway, the project is updated quite frequently and will surely yield some interesting developments.

Another open source project called *Ishmael* [ISH 04] works on an implementation for the *JOnAS J2EE Server* [OBJ 04]. Though the project was registered at the *ObjectWeb* website back in October 2002, it is still considered an alpha release and does not support the current server version. It seems the project's development has nearly stopped. Anyway, in the early development phases of our project 'J2EE Deployment API Implementation' [OLL 04] back in summer 2003, some design decisions were influenced by the Ishmael source code.

### 3   Summary

An approach to enabling reconfiguration of component-based systems at runtime allowing changes of the dependencies among components is presented. We use a meta model which provides description of the system runtime behaviour and a high-level architecture of our reconfiguration manager (

Figure 2). As an implementation platform J2EE Technology [SUN 03] is employed. Currently, we work on an implementation of a J2EE system for runtime reconfiguration of J2EE applications. A special focus of this paper presents our work in progress on implementing a subsystem that enables deployment and redeployment of J2EE modules based on the J2EE deployment API [SEA 03]. Future work includes consideration of simulation methods for predicting the optimal point in time for reconfiguration for a particular reconfiguration request.